\documentclass[runningheads,fleqn]{svmult}

\usepackage{makeidx}   
\usepackage{graphicx}  
\usepackage{subeqnar}  
\usepackage{multicol}  
\usepackage{physmult}  
\makeindex             

\begin{document}

\title*{A model of a pumped continuous atom laser}

\toctitle{A model of a pumped continuous atom laser}
\titlerunning{A pumped atom laser}

\author{Nicholas Robins\inst{1}
\and  Craig Savage\inst{1}
\and Elena Ostrovskaya\inst{2}}

\authorrunning{N. Robins et al.}

\institute{Department of Physics, Australian National University,
ACT 0200, Australia
\and
Optical Sciences Centre, Australian National University,
ACT 0200, Australia}

\maketitle

\section{Introduction}
At the end of the millennium a continuous wave (cw) pumped atom laser remains
to be experimentally demonstrated.  The prototype atom lasers
that have been operated have used rapidly pulsed or continuous output coupling, but
lacked pumping of the trapped atom laser mode \cite{Hagley,bloch}, as 
is needed for a cw laser system. 
Indeed, apart from simple rate equation descriptions, existing
theoretical models do not incorporate both pumping of the laser mode
and propagation of the out-coupled beam.

\begin{figure}[b]
\includegraphics[width=.5\textwidth]{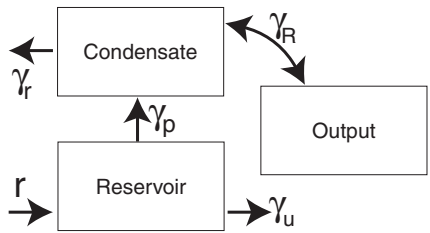}
\caption[]{Schematic representation of the model system.  The symbols 
stand for the coupling, pumping and loss rates and 
are defined in the text.}
\label{modelpict}
\end{figure}

We have used the macroscopic wave-function approximation, or mean field
approximation, which leads to the Gross-Pitaevskii (GP) equation, as the
basis for a complete cw atom laser model.  The macroscopic
wave-function is the order parameter for a Bose-Einstein condensate of 
atoms.   Our model system is schematically illustrated in Fig.  
\ref{modelpict}.  It
includes a pumped reservoir of un-condensed atoms irreversibly
coupled to the atom laser mode \cite{moys}.  The condensed atoms in
this mode, and in the output beam, undergo two-body interactions and
three body recombination.  The atoms are reversibly coupled from the
atom laser mode to the output beam by a Raman process
\cite{Hagley,moys}.  We solve the resulting system of coupled 
differential equations numerically in one spatial
dimension.  The macroscopic wave-function of the output 
atoms is analysed to
determine quantities such as linewidth.
Previous work based on systems of coupled equations of the complex
Ginzburg-Landau type (GP equations with gain and loss)
has either neglected important physics, such as the pumping
\cite{Schneider1,edwards} or has concentrated solely on the condensate,
ignoring the spatio-temporal
structure of the output atom beam \cite{kneer}.

\section{The model}
We consider the trapped atom laser mode, which we refer to as the
\textquotedblleft laser mode\textquotedblright , to be a
one-dimensional condensate in a highly anisotropic, cigar-shaped
harmonic trapping potential.  The atoms in the laser mode and in the
output beam are described by the macroscopic wave-functions
$\Psi_a(x,t)$ and $\Psi_b(x,t)$ respectively.  They experience
two-body repulsive atom-atom interactions and three-body
recombination.  A novel feature of our model is the saturated loss due
to the three-body recombination process \cite{Kagan,burt}.  

To describe pumping we use the phenomenological model of Kneer et.al. 
\cite{kneer} which mimics the pumping mechanism of a conventional optical laser.
The number of atoms in the uncondensed pump reservoir is $N_u(t)$, 
and their spatial distribution is not modeled. 
The reservoir is
fed at a rate $r$, loses atoms at the rate $\gamma_{u} N_u$, and pumps
the laser mode at the rate $\gamma_{p} N_u N_a$,
where $N_{a}(t)=\int^{\infty}_{-\infty}|\Psi_{a}|^2 dx$ is the total
number of atoms in the laser mode.

The laser mode is coupled, by a reversible two-photon Raman
transition, to an untrapped electronic state which is the output atom
laser beam $\Psi_b(x,t)$ \cite{Hagley,moys}.  The Raman transition may
also give a momentum kick to the out-coupled atoms.  In an experiment
the magnitude of this kick depends on the relative geometry of the two
Raman laser beams.  The output atom beam accelerates due to
gravity.  Inside the trap it overlaps,
and hence interacts with, the laser mode atoms.  This interaction has
both a linear contribution, due to the Raman coupling, and a nonlinear
contribution due to the inter-species atom-atom interactions.

Mathematically, our model is described by the following dimensionless 
equations:
\begin{equation}
\label{eq_main}
\begin{array}{l} {\displaystyle
i\frac{\partial \Psi_{a}}{\partial t}=-\frac{1}{2}\frac{\partial^2 
\Psi_{a}}{\partial
x^2}+\frac{1}{2}x^{2}\Psi_{a}+U_{a}\Psi_{a}|\Psi_{a}|^2
+U_{ab}\Psi_{a}|\Psi_{b}|^2}\\*[9pt]
{\displaystyle
 \qquad \quad -i\gamma_{r}\Psi_{a} ( |\Psi_{a}|^4 +|\Psi_{b}|^4 )
 +\gamma_{R} e^{ikx}
\Psi_{b}+\frac{i}{2}\gamma_{p}N_{u}\Psi_{a},}\\*[9pt]
{\displaystyle
i\frac{\partial \Psi_{b}}{\partial t}=-\frac{1}{2}\frac{\partial^2 
\Psi_{b}}{\partial x^2}+Gx\Psi_{b}+U_{b}\Psi_{b}|\Psi_{b}|^2
+U_{ab}\Psi_{b}|\Psi_{a}|^2}\\*[9pt]
{\displaystyle
 \qquad \quad -i\gamma_{r}\Psi_{b} ( |\Psi_{a}|^4 +|\Psi_{b}|^4 )
 +\gamma_{R} e^{-ikx}\Psi_{a},}\\*[9pt]
{\displaystyle
\frac{d N_{u}}{dt}=r-\gamma_{u}N_{u}-\gamma_{p}N_{u}N_{a},}
\end{array}\end{equation}
where $U_{a}$ and $U_{b}$ are the intra- and $U_{ab}$ the 
inter- species two-body
interaction coefficients, $\gamma_{r}$ is the three-body recombination
coefficient, and $\gamma_{R}$ is the Raman coupling coefficient.  

The model is made dimensionless using the characteristic unit of length
$l=(\hbar/\omega m)^{1/2}$ and time, $\tau=\omega^{-1}$, where $\omega$ 
is the trap frequency in the
direction of the weak confinement.  We use $\omega \approx 125$ Hz and 
$m \approx 10^{-26}$ kg. 
The dimensionally correct \cite{edwards} two-body interaction
coefficients, which we assume to be equal, can be written as
$U_{a,b,ab}=4\pi a_{s}/l $ where $a_{s}$ is the s-wave scattering
length for a specific process.  There is some arbitrariness in how the $U_{a,b,ab}$ relate to
their three dimensional counterparts.  The scaling we have chosen
gives a realistic condensate size \cite{edwards}.  Similar reasoning
applies to the choice of the three-body recombination rate, $\gamma_{r}$.  We have
set it to give feasible lifetimes for the laser mode condensate:
it is typically between $10^{-7}$ and $10^{-9}$ in our simulations. 
In accordance with experimentally reasonable values, we allow the
dimensionless Raman coefficient,
$\gamma_{R}=\Omega_{1}\Omega_{2}/(\omega \Delta)$, to range up to
$10^{4}$, where $\Omega_{1,2}$ are the Rabi frequencies of the Raman
lasers and $\Delta$ is the laser detuning \cite{moys}, and we take the
dimensionless Raman momentum kick to lie in the range $0 \leq k \leq
100$.  The dimensionless gravitational coefficient $G=g (m\omega^{-3}
\hbar^{-1})^{1/2}$ varies in the range $0\leq G \leq 68$, with $G=68$ 
corresponding to $g=9.8$ ms$^{-2}$.  Physically,
we may think of the output atomic beam  propagating in a tilted atom waveguide
\cite{marksteiner}, with the tilt angle determining the value of $G$.

\section{Rate equations}
Given the complexity of the model equations (\ref{eq_main}), a numerical method
 is the only feasible method of solution.
However, some understanding of the dynamics of the system can be gained by deriving 
approximate rate equations for the populations of the reservoir, 
condensate and output fractions \cite{nick}. 
We proceed by projecting the
condensate and output mean fields onto the stationary state modes
$\Phi_a(x)$ and $\Phi_b(x)$ defined by:
\begin{equation}\label{anz}
\Psi_{a}(x,t)=a(t)\Phi_{a}(x)\qquad 
\Psi_{b}(x,t)=b(t)\Phi_{b}(x)e^{-ikx},
\end{equation}
where $\Phi_{a,b}$ satisfy the time independent, lossless, and decoupled form of
the equations (\ref{eq_main}):
\begin{equation}
\label{eq_3}
\begin{array}{l} {\displaystyle
\frac{1}{2}\frac{\partial^2 
\Phi_{a}}{\partial
x^2}=-\mu_{a}\Phi_{a}+\frac{1}{2}x^{2}\Phi_{a}+U_{a}\Phi_{a}|\Phi_{a}|^{2},}\\*[9pt]
{\displaystyle
\frac{1}{2}\frac{\partial^2 \Phi_{b}}
 {\partial x^2}=-\mu_{b}\Phi_{b}+Gx\Phi_{b}+U_{b}\Phi_{b}|\Phi_{b}|^{2}.}
\end{array}
\end{equation}
We are free to choose the magnitude of the chemical potentials $\mu_{a}$ and $\mu_{b}$.
\begin{figure}
\includegraphics[width=\textwidth]{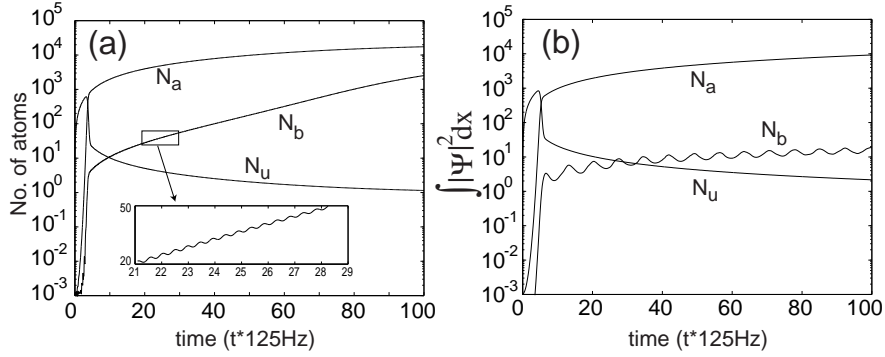}
\caption[]{Typical population dynamics without three-body recombination.  (\textbf{a}) From the rate
equations (\ref{pop}) showing the growth of the laser mode.  We have tested that this growth
 continues in the long time limit ($\tau > 600$). Parameters
are $r=200$, $\gamma_{p}=\gamma_{u}=0.01$, $\gamma_{R}=0.5$,
$U_{a}=U_{b}=0.02$, $U_{ab}=0.01$, $\delta \mu=0.02$, $N_{a}(t=0)=0.001$,
$N_{b}(t=0)=10$ and $\gamma_{r}=0$. Values of the spatial integrals are as given in the text.  The inset
shows the Rabi-type oscillations, present in the condensate and output fields.
 (\textbf{b}) From the full GP equations (\ref{eq_main}). 
	Parameters are the same as for part (a) and G=68, $k=5$. 
 The GP simulations predict a much smaller number of atoms in the output beam, $N_{b}$,
than the rate equations, because there is an effective
loss of atoms due to the boundary absorber at the edge of the numerical grid.}
\label{ratevgp}
\end{figure}
In order to simplify the analysis we set the momentum kick, $k$, 
equal to zero.  By doing so, we neglect the main effect the momentum kick 
has on the dynamics of the system, that is, the change in the spatial overlap 
between the condensate and the out-coupled field.

The first equation in (\ref{eq_3}) describes the eigenfunction of a trap with
chemical potential $\mu_a$.  The
equation for the output mode describes a falling wave and can be
solved numerically or perturbatively, using a superposition of Airy
functions in the linear limit \cite{Schneider2}.  We now substitute the
ansatz (\ref{anz}) into the original system (\ref{eq_main}), and 
use the equations (\ref{eq_3}) to simplify the result.  We then 
integrate out the spatial dependence in the equations and
separate the time dependence of the modal amplitudes  
as $a=n_{a}(t)e^{i\theta_{a}(t)}$ and $b=n_{b}(t)e^{i\theta_{b}(t)}$.
By noting that the relationship between the modal amplitudes and the population 
numbers is given by $N_{a}=I_{6}\delta n_{a}^{2}$ and
$N_{b}=I_{5}\delta n_{b}^{2}$, we arrive at the following system of equations for the real 
population numbers $N_{a,b}$, $N_u$, and phase difference
$\theta=\theta_b-\theta_a$:
\begin{equation}
\label{pop}
\begin{array}{l} {\displaystyle
\frac{d N_{a}}{d t}=-(\alpha N_{a}^{3}+\beta N_{a}N_{b}^{2})+\gamma_{p}N_{u}N_{a}+
2\gamma_{R}\sqrt{I_{5}I_{6}N_{a}N_{b}}\sin(\theta)},\\*[9pt]
{\displaystyle
\frac{d N_{b}}{d t}=-(\beta N_{b}^{3}+\alpha N_{b}N_{a}^{2})
-2\gamma_{R}\sqrt{I_{5}I_{6}N_{a}N_{b}}\sin(\theta)},\\*[9pt]
{\displaystyle
\frac{d \theta}{d t}=\delta \mu+\delta_{NL}+
\frac{\gamma_{R}\sqrt{I_{5}I_{6}}}{\sqrt{N_{a}N_{b}}}\cos(\theta)(N_{b}-N_{a})}\\*[9pt]
{\displaystyle
\frac{d N_{u}}{dt}=r-\gamma_{u}N_{u}-\gamma_{p}N_{u}N_{a}},
\end{array}
\end{equation}
where $\delta \mu=(\mu_{a}-\mu_{b})$, $\alpha=(2\gamma_{r}I_{3})/(\delta I_{6})^{2}$,
$\beta=(2\gamma_{r}I_{4})/(\delta I_{5})^{2}$, and
$\delta_{NL}=U_{b}(1-N_{b}/(\delta I_{5}))I_{2}+U_{a}(N_{a}/(\delta I_{6})-1)I_{1}+U_{ab}((N_{a}I_{1})/(\delta I_{6})-(N_{b}I_{2})/(\delta I_{5}))$.

The constant coefficients $I_{s}$, where $s=1,2,3,4,5,6$, and $\delta$ are determined by 
spatial overlap integrals as follows:
\begin{equation}
    \begin{array}{l}{\displaystyle
I_{1}=\delta^{-1} \int \Phi_{a}\Phi_{b}|\Phi_{a}|^{2}dx, \qquad \quad I_{2}=\delta^{-1} \int 
\Phi_{a}\Phi_{b}|\Phi_{b}|^{2}dx,}\\*[9pt]
{\displaystyle
I_{3}=\delta^{-1} \int 
\Phi_{a}\Phi_{b}|\Phi_{a}|^{4}dx, \qquad \quad I_{4}=\delta^{-1} \int 
\Phi_{a}\Phi_{b}|\Phi_{b}|^{4}dx, }\\*[9pt]
{\displaystyle
I_{5}=\delta^{-1} \int |\Phi_{b}|^{2}dx, \qquad I_{6}=\delta^{-1} \int 
|\Phi_{a}|^{2}dx, \qquad \delta=\int \Phi_{a}\Phi_{b}dx.}
\end{array}\end{equation}
For realistic parameters, and using the stationary solutions of 
Eq. (\ref{eq_3}) we numerically calculated these integrals to have the approximate
 values $\delta=233$, $I_{1}=875$, $I_{2}=1.45$, $I_{3}=8.39\times10^{5}$, 
$I_{4}=2.79$, $I_{5}=0.13$ and 
$I_{6}=41.3$.

From the rate equations it is seen that the populations of the 
condensate and the output mode undergo nonlinear Rabi-type 
oscillations \cite{ballagh} with the effective frequency determined, 
in part, by the strength of the interactions through 
$\delta_{NL}(U_a,U_b,U_{ab})$, and the Raman coupling $\gamma_R$. In the 
presence of pumping and without the three-body loss terms,
 the condensate fraction acquires a 
fast growing component which prevents the system from reaching a 
steady state defined by $dN_{a,b}/dt=0$.    A typical solution to 
the system (\ref{pop}) is shown in Fig. \ref{ratevgp}(a).  
An important feature of this 
regime is the unbounded growth of the laser mode in the absence of three-body recombination, 
which would not be the case for a realistic atom laser.
 This emphasises the need to include three-body
recombination in atom laser models, as we do in subsequent sections.
 The qualitative predictions of the rate 
equation model are confirmed by direct integration of the system 
Eq. (\ref{eq_main}), with the results shown in Fig. \ref{ratevgp}(b). 
It is important to note that the values of the spatial integrals, 
 $I_{s}$ and $\delta$, play a large part in the time 
evolution of the rate equations.
\begin{figure}
\includegraphics[width=\textwidth]{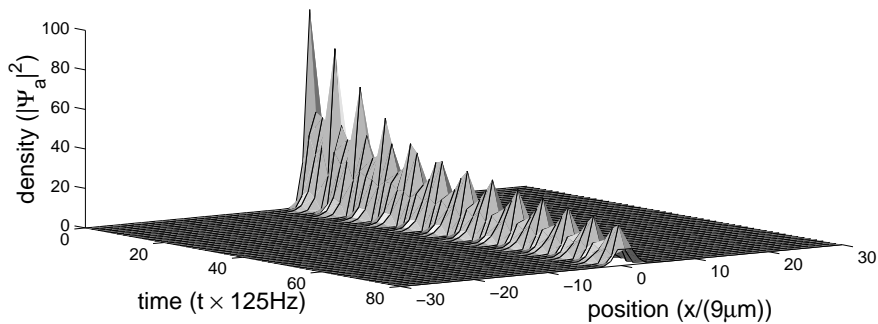}
\caption[]{The time evolution of an unpumped condensate.  Parameters are 
$\gamma_{R}=0.5$, $\gamma_{r}=10^{-7}$, $U_{a}=U_{b}=0.02$, $U_{ab}=0.01$, $G=12$, $k=5$, and $N_{a}(t=0)=140$}
\label{decay}
\end{figure}
\section{Numerical results}

With this initial understanding of the behavior of the atom laser to guide us, we
proceed to investigate the numerical solution of our full model equations
(\ref{eq_main}).  We use a split-step method \cite{taha} to analyse
the spatio-temporal behavior of the laser mode and the output beam.  An
absorbing boundary is used at the edges of the spatial grid.  We have
ensured that our results are insensitive to the size of the spatial
grid and time-step.

Our model is at its simplest with the pump mechanism turned off, so that 
the out-coupling depletes the atom laser mode.  The laser mode
population changes rapidly at the beginning of the
first Rabi oscillation, as can be seen in Fig.\ref{decay}, so the approximation of a
stationary mode may be false even on short time scales
\cite{naraschewski2}.  One of the results of this paper is that it might be
 more realistic to make this 
approximation on the assumption of a steady state achieved through a 
balance of pumping and loss due to three body recombination.

\begin{figure}
\includegraphics[width=.7\textwidth]{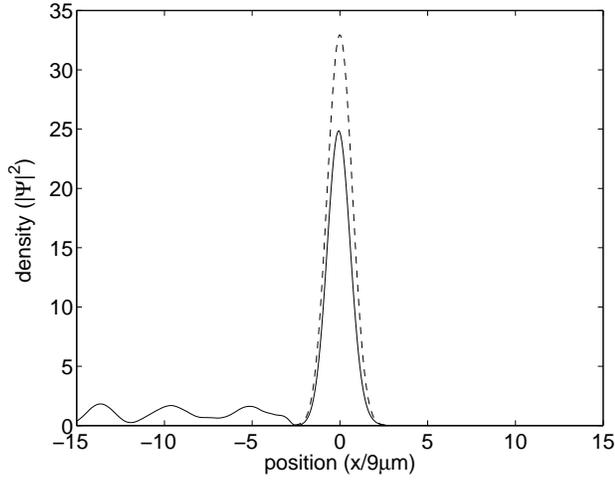}
\caption[]{The bound state of the laser mode (\emph{dashed
line}) and the output beam (\emph{solid line}) at $t=200$.  The laser mode
 is multiplied by 0.003.   The output beam, for $x < -2.5$ only, is multiplied by 
a factor of 10.  Parameters are $r=200$,
$\gamma_{R}=0.5$, $\gamma_{r}=0$, $U_{a}=U_{b}=U_{ab}=0$, $G=0$, $k=5$,
$N_{a}(t=0)=0.001$, $N_{b}(t=0)=10$.}
\label{trappedstate}
\end{figure}

\begin{figure}
\includegraphics[width=\textwidth]{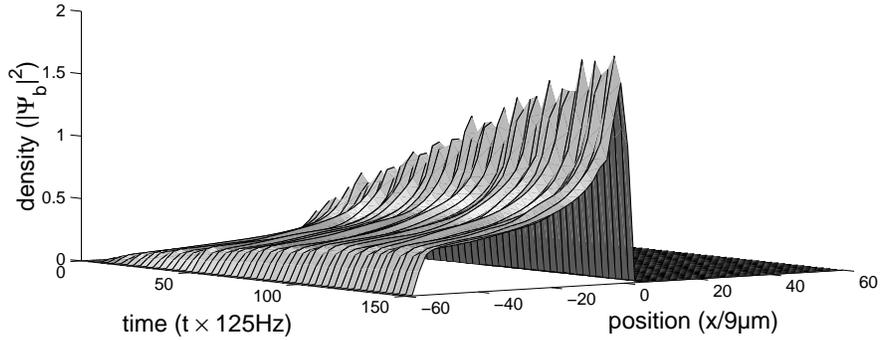}
\caption[]{Atom laser beam density as a function of position and time,
with no three-body recombination, showing the steady growth of the
output.  The beam propagates in the negative $x$ direction from the 
condensate centered at $x=0$. 
Parameters are $r=200$, $\gamma_{R}=0.5$, $\gamma_{u}=0.1$,
$\gamma_{p}=0.1$, $\gamma_{r}=0$, $U_{a}=U_{b}=0.02$, $U_{ab}=0.01$, $G=68$, $k=5$,
$N_{a}(t=0)=0.001$, $N_{b}(t=0)=10$.}
\label{basis}
\end{figure}

We next consider the pumped system.  Without the effects of gravity
and interactions we find the bound state predicted by Hope et al. 
\cite{hope3}.  This is produced by the Raman coupling.  A
fraction of the output field remains localised around the laser mode,
see Fig.\ref{trappedstate}, and hence the population in the laser mode
increases indefinitely due to the pumping.  Hope et al. found that
including gravity or two-body interactions could destroy this bound state by
introducing a repulsive effective potential acting on the output mode. 
A large fraction of the atoms then quickly leave the interaction
region.  However we find that when both gravity \emph{and} interactions act together they tend
to recreate the exponentially growing output state.  The population
dynamics of the laser mode and the output is approximately described
by Eqs.(\ref{pop}) in this case.  The output beam's spatio-temporal structure, in the absence of
three-body recombination, is shown in Fig.\ref{basis}.
\begin{figure}
\includegraphics[width=\textwidth]{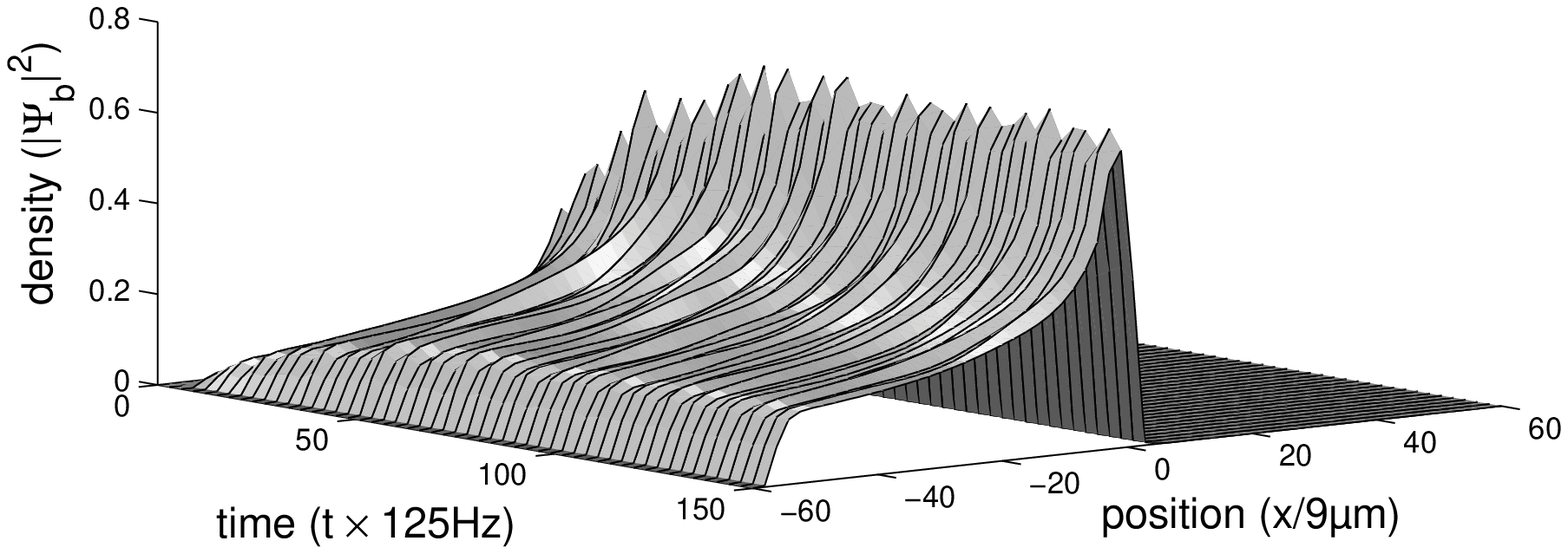}
\caption[]{Atom laser beam density as a function of position and time, 
with three-body recombination included $\gamma_{r}=10^{-7}$, showing a 
quasi-steady state.  Parameters are otherwise the same 
as for Fig. \ref{basis}.}
\label{nice}
\end{figure}

\begin{figure}
\includegraphics[width=\textwidth]{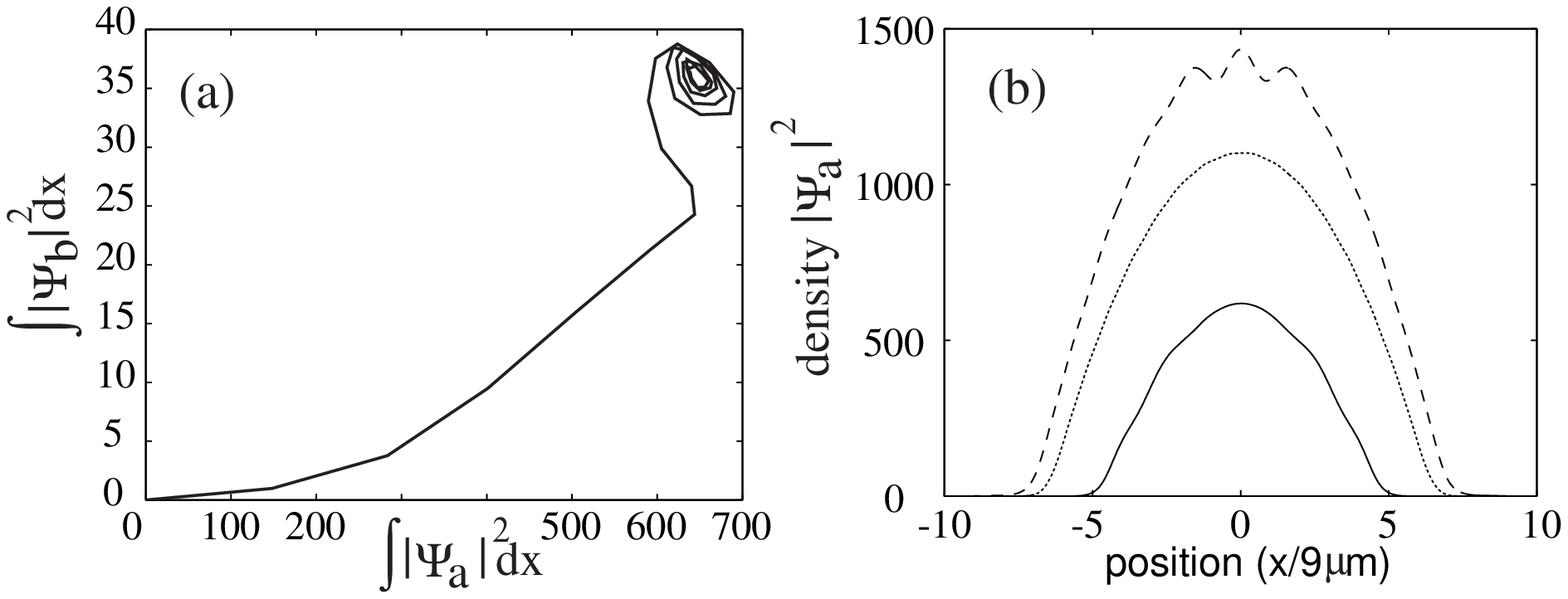}
\caption[]{Steady state operation of the system with three-body
recombination.  (\textbf{a}) Atom number in the output versus atom
number in the laser mode.  Parameters are $r=100$, $\gamma_{R}=0.5$,
$\gamma_{u}=0.1$, $\gamma_{p}=0.1$, $\gamma_{r}=10^{-5}$,
$U_{a}=U_{b}=0.02$, $U_{ab}=0.01$, $G=12$, $k=5$, $N_{a}(t=0)=0.001$,
$N_{b}(t=0)=100$.  
(\textbf{b}) The quasi-steady state density profile of the laser mode
at $t=150$ for pumping of $r=200$ (\emph{solid}) and $r=1600$
(\emph{dotted})  Other parameters are  $\gamma_{R}=0.5$, $\gamma_{u}=0.1$,
$\gamma_{p}=0.1$, $\gamma_{r}=10^{-7}$, $U_{a}=U_{b}=0.02$, $U_{ab}=0.01$, $G=68$, $k=5$,
$N_{a}(t=0)=0.001$, $N_{b}(t=0)=10$.  By contrast the laser mode at $t=150$ for pumping of
$r=200$ (\emph{dashed}), with \emph{no} three body recombination, is
continuing to grow and change shape.}
\label{steady}
\end{figure}

Adding three body recombination destroys the bound state as can be 
seen in Fig. \ref{nice} and Fig. \ref{steady}(a).   This occurs because the
loss rate increases as the square of the atomic density.  In addition, the 
three-body recombination effect supresses the atomic density noise 
in the output field, as well as supressing the collective excitations of 
the condensate, see Fig. \ref{steady}(b).  Remarkably, the three body 
recombination process assists in creating a steady state, both in the 
output and laser modes, as can be seen in Fig. \ref{steady}(a). 

\section{Linewidth narrowing}
One of the characteristic properties of a continuous wave optical
laser is the dramatic reduction in linewidth that occurs above
threshold.  The ultimate quantum mechanical limit to the linewidth is
given by the Schawlow-Townes formula \cite{schawlow,wiseman1}.  Graham
has proposed a related limit for the atom laser due to interaction
with the thermal reservoir of uncondensed atoms \cite{graham}.  Our
model does not incorporate the many-body quantum physics that
determines the ultimate atom laser linewidth.  However, in practice we
might expect that the linewidth is not determined by fundamental
factors, but rather by the dynamics of the particular system; for
example, in the optical case by relaxation oscillations
\cite{siegman}.  Our model is suitable for determining the linewidth
due to this type of dynamical effect.

\begin{figure}
\includegraphics[width=.7\textwidth]{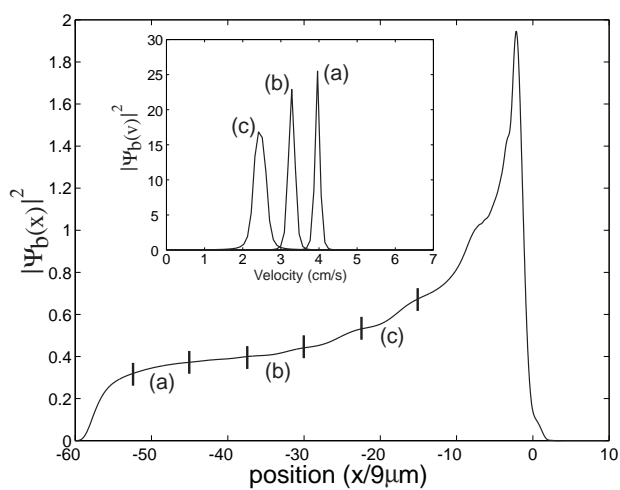}
\caption[]{The density of the output beam as a function of position. 
The inset shows the velocity linewidth over three regions of the atom
laser output. These regions are labeled (a), (b), and (c) and are 
indicated by the bars on the main curve.  Their width is about 7.5 
dimensionless units, or 68 $\mu$m. The number of atoms in each 
region is approximately the dimensionless density times the length; 
e.g. $7.5 \times 0.4=3$ for region (b). Parameters are as in 
Fig.\ref{steady}(a).}
\label{linewidth}
\end{figure}

To determine the linewidth of the output we have calculated the
instantaneous spatial Fourier transforms of particular regions of the
output field.  The results are shown in the inset to
Fig.\ref{linewidth}.  The atoms are accelerated by gravity according
to $v=\sqrt{v_{0}^{2}+2gx}$ and hence, the velocity linewidth narrows
as the atoms fall away from the trap \cite{wiseman2}.  We are currently 
investigating the effect of the pump rate, $r$, on the output linewidth of 
our model, in both the spatial and temporal domains.  

\section{Conclusion}
We have presented a model of an atom laser in which we have endeavored
to include the important physics of an experimental atom laser system. 
The resulting model exhibits a rich and complex dynamics.  A novel
aspect of our model is the inclusion of three-body recombination, which
we believe plays an important part in achieving steady state laser
operation.  We demonstrated that the complex 
interplay of effects in our model can lead to a steady output with low atomic 
density fluctuations.  The output of our atom laser is locally monochromatic, its
linewidth dominated by gravitational acceleration.  We are in the process of
 generalising this one dimensional model to three dimensions.

We acknowledge continuing discussions with J. Hope, and thank G. Moy 
for his input.

\end{document}